\documentclass[5p, number,times]{elsarticle}
\pdfoutput=1

\usepackage{amssymb}
\usepackage{amsmath}
\usepackage{graphicx}
\usepackage{natbib}
\usepackage{threeparttable}

\journal{Ultramicroscopy}
\begin{document}
\begin{frontmatter}
\title{Comparison of optimal performance at 300~keV of three direct electron detectors 
for use in low dose electron microscopy}

\author[MRC]{G.~McMullan \corref{cor1}}
\ead{gm2@mrc-lmb.cam.ac.uk}
\cortext[cor1]{Corresponding author}

\author[MRC]{A.R.~Faruqi}
\author[BB]{D.~Clare}
\author[MRC]{R.~Henderson}
\address[MRC]{MRC Laboratory of Molecular Biology, Francis Crick Avenue, Cambridge CB2 0QH.}
\address[BB]{Crystallography and Institute of Structural and Molecular Biology, Birkbeck College, University of London, Malet Street, London WC1E 7HX.}

\begin{abstract}
Low dose electron imaging applications such as  electron cyro-microscopy are  
now benefitting from the improved performance and flexibility of 
recently introduced electron imaging detectors in which electrons are
directly incident on backthinned CMOS sensors. There are currently 
three commercially available detectors of this type: the 
Direct Electron DE-20, the FEI Falcon~II and the Gatan K2 Summit.  These
have different characteristics and so it is important to compare 
their imaging properties carefully with a view to optimising how each is used. 
Results at 300 keV for both the modulation transfer function (MTF) and 
detective quantum efficiency (DQE) are presented.  Of these, the DQE is the most important
in the study of radiation sensitive samples where detector performance
is crucial.  We find that all three detectors have a better DQE than film.
The K2 Summit has the best DQE at low spatial frequencies but with increasing 
spatial frequency its DQE falls below that of the Falcon II.

\end{abstract}


\end{frontmatter}

\section{Introduction}
Electron microscope images were originally recorded on photographic
film and more recently electronically using detectors based on 
phosphor/fibre-optic CCD technology.  These work well for 
electron energies in the 80-120 keV range but at higher electron energies
their performance drops. Higher electron energies are necessary with 
thicker samples and advantageous in looking at insulators, such as 
ice embedded biological samples, due to the reduced sensitivity to sample charging.
The shorter wavelength also results in improved 
electron optics and simpler interpretation of the resulting images.

The decrease in imaging performance of traditional detectors at higher electron energies
can be traced to reduction in the interaction cross-section with increased
energy.  Higher energy electrons deposit a lower, and more variable, amount of energy 
at their initial point of incidence. Their subsequent path has a far greater range
leading to the appearance of tracks in the detector and contributions from where 
electrons backscatter from either deeper within the 
substrate of the detector or from the surrounding housing.
The lower, and more variable,
initial signal combined with the addition of backscattering events that contribute
to the noise results in a lower signal to noise even near zero spatial frequency. At
higher spatial frequencies the performance is degraded by 
the stochastic nature of electron trajectories and the fact that
the rate of energy loss by an electron increases as the electron slows down.

Detector performance is of particular importance in the study
of radiation sensitive samples such as in electron microscopy (cryoEM), 
where the signal to noise ratio in images is inherently poor
due to the limited number of electrons that can be used before
radiation damage is too great.
The amount of additional noise
added by a detector is measured by its 
detective quantum efficiency (DQE) which is
defined\cite{DaintyShaw} as the square of the ratio of the 
output signal to noise, $\mathrm{SNR}_\mathrm{o}$, to
that of the input, $\mathrm{SNR}_\mathrm{i}$, i.e., 
\begin{equation}
\mathrm{DQE} = \mathrm{SNR}_\mathrm{o}^2 / \mathrm{SNR}_\mathrm{i}^2.
\end{equation}
Ideally a detector would not add any noise and so have a DQE of 1 but 
all real detectors have values less than 1.

Direct detection of electrons using backthinned monolithic active pixel 
sensors (MAPS) has emerged as the most promising technology with which to
produce detectors  with high DQE at higher incident electron 
energies\cite{Faruqi2005,Milazzo2005152,McMullan20091126}.
MAPS detectors are fabricated in silicon using industry standard 
CMOS imaging technology that enables the manufacture of uniform large 
format ($\ge 4k \times  4k$) pixel sensors. Their
potential for use as high DQE detectors is reflected in the high signal 
to noise with
which they are capable of detecting individual incident 300~keV electrons.
They are susceptible to radiation damage and despite an increase in
radiation hardness with smaller dimension fabrication technology, any
practical detector must make use of radiation-hard design techniques.

The range of a 300 keV electron in silicon can exceed $300\,\mu$m and 
so it is not practical to limit the 
signal from an  individual incident electron to a single pixel. MAPS detectors 
can however have most of their support matrix removed so that a functioning detector 
can consist of only a thin membrane ($\leq 50\, \mu$m) 
through which incident 300 keV electrons can easily pass.  In order to maximise the benefit 
of this process it is also important to mount a detector carefully to prevent
transmitted electrons from scattering back into the detector from the camera housing.

MAPS detectors are capable of high readout
speeds and this can be used to ameliorate the effects of radiation damage by 
limiting the  contribution in any frame from increased leakage current 
associated with radiation damage.   The combination of high DQE and
high readout speed also gives greater flexibility in imaging. For example, images
can be recorded as dose-fractionated movies from which 
the optimal exposure 
can be selected during image processing, long after the 
specimen has been removed from the microscope.  The combination of high 
sensitivity and readout speeds makes it possible to use a counting mode in which a final image
is reconstructed from processed sub-images of individual electron events\cite{McMullan20091411}. 
This enables the intrinsic variability in the signal left by an incident
high energy electron to be removed and so achieve a higher
DQE, at least at low spatial frequency. It is also possible to infer the initial point of 
incidence of an electron to sub-pixel resolution (super-resolution mode) and so 
obtain information beyond the traditional Nyquist frequency limit.

In this paper we present MTF and DQE measurements of 
the three currently available backthinned MAPS detectors that
offer improvements over photographic film in terms of 
DQE at 300 keV namely: the Direct Electron DE-20 \footnote{Direct Electron, LP, www.directelectron.com}; 
the FEI Falcon~II \footnote{FEI, www.fei.com};  
and the Gatan K2 Summit \footnote{Gatan, Inc., www.gatan.com}.
It is possible to record dose-fractioned 
movies with all three detectors and in principle 
operate any of the detectors in a counting mode.  In this paper
only counting mode results obtained using the K2~Summit will be presented.

\section{Methods}

The Falcon~II and K2 Summit detectors were both installed on a FEI Titan 
Krios at the MRC Laboratory of Molecular Biology (MRC-LMB) in Cambridge while the
DE-20 detector was installed on a FEI Polara G2 at Birkbeck College in London.
Both microscopes were fitted with Gatan energy filters 
(GIF Quantum on the Titan Krios and GIF 2002 on the G2 Polara).
The Falcon~II and DE-20 detectors were both positioned before 
their respective energy filters while the K2 Summit was positioned after the energy filter. 
The energy filter was operated without an energy slit and
carefully tuned before any measurements were taken.

In order to compare the detectors
on different microscopes accurately the exposure meter on each microscope was first
calibrated.  To do this the exposure meter readings were compared at series of 
different beam currents obtained by altering the spot size.
The beam current was measured using the 
drift tube of the energy filter as a Faraday cup.
The same SEM (scanning electron microscopy) probe current 
meter\footnote{SEM Probe Current Meter, part No. 087-001, www.deben.co.uk} was 
used to measure the beam currents on both microscopes.
The accuracy of this meter was confirmed to better than 1\% in the range of interest 
using a calibrated laboratory voltage standard and combinations of high precision 100 M$\Omega$ resistors. 
For a given spot size a small diameter beam was
positioned so that with the energy filter set to 300 keV the beam could be seen to pass 
entirely through the energy filter.   The beam current was then measured by
setting the energy filter to zero energy so that the 
entire beam hit the drift tube.
The corresponding microscope exposure 
meter reading was obtained by lowering the flu-screen.
As the beam current can vary with
changes in either the condenser or objective lens strengths, care was taken not
to alter the beam settings during a measurement.
After each current measurement the energy filter was set back to 300 keV in order
to verify that the entire beam once again passed entirely 
through the energy filter.
On both microscopes the exposure meter readings obtained this way 
could accurately be described as a linear function of 
the corresponding probe current meter readings, though with different slopes and offsets.
The calibrations of both the Krios and Polara are given in supplementary
information.  Allowing for the uncertainty in the exposure meter reading the
total current in a beam could be measured to within 3\%.

The number of electrons incident per second on a pixel was calculated
from a measured current in a defined circular beam contained entirely on 
the detector. To minimise Fresnel effects at the edge of the beam, a
selected area aperture was used to define the beam on the detector and
the lowest possible magnification used.
The microscope exposure 
meter on the Krios microscope did not register currents below $42\,$pA. 
To measure lower currents a beam with greater than $42\,$pA 
was first set and the required, incident rate for electrons on the
detector obtained by increasing the microscope magnification.
Unlike the case with changes in either the objective or condenser lens
the measured beam current is not sensitive to changes in magnification resulting
from the diffraction, intermediate or projector lenses, provided the entire beam 
remains on the flu-screen (which provides the input to the exposure meter).
The relative magnifications between the different microscope settings
was calibrated using images at the different magnifications centred 
on the same area of a sample.

Backthinning a detector by itself does not guarantee improved performance
from a detector. In particular, backscattering 
from the silicon substrate of a detector may simply be replaced
by backscatter from the aluminium alloy that
typically lines a camera housing. As the housing is further away
the backscattering contribution will be moved to lower 
spatial frequency.
The simplest and most effective way to reduce the amount of 
backscatter is to increase the distance between the 
backthinned detector and its housing.
For large detectors such as the Falcon~II it is not easy to
find sufficient space under the detector and additional steps such as 
replacing any aluminium by lighter elements such as beryllium, boron 
or carbon are needed.
The amount of backscatter from the housing  can be measured by taking a series of images
in which the edge of a large selected area aperture is scanned across the detector. 
Electrons passing through
a detector undergo many collisions and so by the time they leave the detector their
probability of being backscattered to the detector by the housing can be
taken to be uniform. 
The presence of  backscatter
shows up in both the illuminated and shadow areas  
as an additional  contribution that is proportional to the area of the detector being illuminated.
With the DE-20 this contribution was found to be
5\% while for  both the K2 and Falcon~II detectors it was
less than 2\% .

The DQE of detectors is in general dependent on the dose rate and for this work 
we have attempted to use a dose rate that was optimised for each detector.
For the K2 Summit this meant using as few electrons per frame as possible.
A value of  1.1 e/pixel/sec, or 1 electron per every 360 pixels
in an individual frame, was used. For both the DE-20 and FEI Falcon~II the dose rate 
was chosen so that the peak in the histogram from an image of an individual frame 
was positioned at approximation 1/3 of the detector's dynamic range. This
criterion meant that the DE-20 and Falcon~II were operated with 4 and 3 
electrons per pixel per frame, respectively. As the DE-20 was operated at 25~fps
it was therefore used at twice the 
dose rate (expressed as electrons/pixel/sec) of the Falcon~II operating at 18~fps.
In practice the high signal to noise seen in both the 
DE-20 and Falcon II means that much lower dose rates can also be used 
without significantly degrading the DQE.

The MTF and DQE were measured using the
procedure described in \cite{McMullan20091126}.
The MTF was measured using the shadow image of a platinum rod.
In the Polara this consisted of $2\,$mm diameter rod inserted 
at the pointer position. This was not possible with the 
Krios but as the microscope was fitted with a film mechanism 
a modified film holder was  used to support a $1\,$mm rod. This  
was positioned manually using a syringe to pressurise the
film insertion mechanism.  
The accuracy of the MTF obtained from the 
shadow image of a sharp edge depends on the quality of the edge.
Only straight, blemish free sections from the images were used.
For measuring the MTF of the K2~Summit extra care was needed due to its
intrinsically high MTF, small pixel size and the additional x6 magnification
from the energy filter. The quality of an edge was
verified by examining the difference between the original image
and a simulated edge image blurred by the fitted MTF. As in \cite{McMullan20091126}
a sum of Gaussian functions is used as a convenient analytical fit. 
This is known to be an ill-posed and the validity of any fit
was always checked with results from direct numerical 
differentiation of the measured edge spread function.

The MTF of the Falcon II was also calculated using the noise power
method\cite{DaintyShaw}. In this the signal from
an incident electron is described by
a circularly symmetric point spread function, PSF, and the MTF
obtained from the Fourier transform of this PSF. In this 
work the PSF is expanded as a normalised sum of Gaussian functions
in which the weights and length parameters are obtained
by fitting to the measured noise power spectra (see Appendix~A).
The noise power method typically over estimates the high
frequency MTF, since the 
noise power spectra reflects the stochastic 
response to individual electrons rather than the averaged
response as measured by the MTF \cite{MeyerKirkland1998, Boothroyd201318}.

The gain of a counting detector, such as the K2 Summit,
decreases as the probability of two or more electrons being 
recorded in or around a pixel increases (see Appendix~B). 
The resulting non-linearity will in general 
need to be corrected for, especially in high contrast images
such as that of the sharp edge used to measure the MTF. 
To avoid this complication, the MTF of the K2~Summit was 
measured with a very low dose rate ($\sim$~1~e/pixel/sec) that required
using long exposures ($\sim$~300 sec).

\begin{table}
\centering
\caption{Physical properties of the detectors}
\smallskip
\begin{threeparttable}
\begin{tabular}{rcrr}
\hline
Detector & Sensor size  & Pixel size  & Readout speed  \\
  & ($\mu$m) & (fps)  \\
\hline
DE-20   &     5120 x 3840 & 6.4   &   25\tnote{a} \\
Falcon-II   & 4096 x 4096 & 14.0  & 18   \\
K2 Summit  &  3838 x 3710 & 5.0  &  400  \\
\end{tabular}
\begin{tablenotes}
\item[a]{\footnotesize  The DE-20 is capable of operating up to 32.5 fps}
\end{tablenotes}
\label{table:PROPERTIES}
\end{threeparttable}
\end{table}


\section{Results}

\begin{figure*}[t]
\begin{center}
\includegraphics[width=14cm]{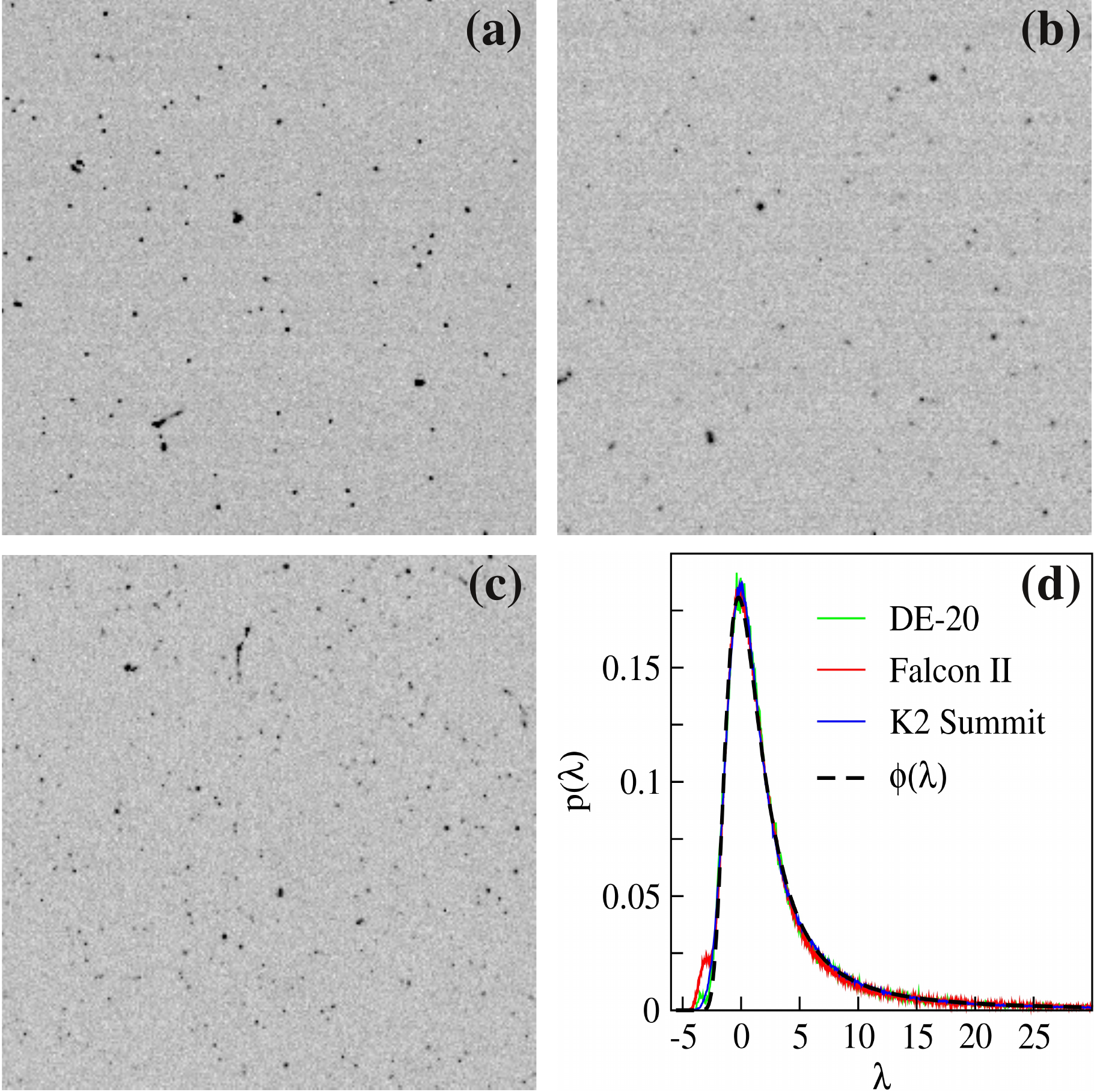}
\end{center}
\caption{
Randomly chosen 256x256 areas from single frames 
showing individual 300 keV electron events as recorded on 
the (a) DE-20, (b) Falcon II and (c) K2 Summit detectors.
The images are normalised so that the RMS background noise 
has  a value of 1. In (d) the 
Landau distribution, $\phi(\lambda)$, is compared  with
the measured probability distributions for
events as a function of integrated event signal, $\Delta$, in
the three detectors. The measured distributions have
been scaled by the fitted width parameter, $\xi$, and plotted using 
$\lambda = [\Delta - (\Delta_{\mathrm{mp}} - \xi\lambda_0)] /\xi$ in 
which $\lambda_0 = - 0.2228$ is position of the maximum of 
$\phi(\lambda)$,  $\Delta_{\mathrm{mp}}$ is the position of the most probable
value.
The measured ratios of $\Delta_{\mathrm{mp}}$ to the RMS background noise for the
DE-20, Falcon~II and K2~Summit are 49.6, 30.6 and 19.6 respectively, while
the corresponding ratios
of $\Delta_{\mathrm{mp}}$ to $\xi$ are  4.8, 5.3, and 5.4. 
}
\label{fig:EVENTS}
\end{figure*}

The physical properties of the three detectors are summarised in Table~\ref{table:PROPERTIES}.
The high sensitivity of MAPS detectors to individual 300 keV electrons 
is illustrated in Fig.~\ref{fig:EVENTS}(a,b,c). These show
images of single electron events as recorded in single frames on the detectors.
Single frame output is not currently supported
on the K2~Summit and in order to obtain this image 
the data-stream from the camera was intercepted and decoded. 
The actual frames and  the parts of them that are shown were chosen randomly.
In order to compare the detectors the signals have
been scaled so that the RMS readout noise in each pixel is 1. The intrinsic variability 
in the events can be clearly seen as well as the presence of the occasional 
electron track. 

In high energy physics the intrinsic variability of the energy 
loss by a charged particle passing through a thin absorber 
is known as straggling and the distribution of energy 
loss usually fitted to a Landau distribution, $\phi(\lambda)$  \cite{Landau1944}.
In Fig.~\ref{fig:EVENTS}d the measured probability distributions for 
integrated signal of individual events as a function of signal  are
compared with the Landau distribution. 
The distributions were calculated from
images in which individual incident electrons could easily be resolved.
A threshold of 4 times the average readout noise was used to identify a seed
pixel of an event.  Having identified a seed pixel 
all contiguous pixels also above the threshold were used to define an event and the
total signal, $\Delta$, for an event obtained 
by summing the contributions from pixels both in the event and form within
a radius of 2 pixels around the event. The probability distribution, $p(\Delta)$,
for events was then fitted to a scaled Landau distribution   
$\phi( \lambda )/\xi$ in which 
$\lambda =  [\Delta - (\Delta_\mathrm{mp} - \xi \lambda_0)]/\xi$,
$\lambda_0 = - 0.2228$  is the position of the maximum of $\phi(\lambda)$,
$\xi$ is a fitted width parameter and 
$\Delta_\mathrm{mp}$ 
the position of the most probable value of $p(\Delta)$.
As can be seen in 
Fig.~\ref{fig:EVENTS}d, the measured distributions 
fit the functional form of Landau distribution very well.
The small pedestal in the measured distributions  visible
at low $\Delta$ is due to the erroneous inclusion of 
noise events and can be removed using a higher threshold.
The absolute scale of $\Delta$ in terms of energy was not calibrated but
the ratio of $\Delta_\mathrm{mp}$  to the noise gives a measure
of the signal to noise in the detector. 
The values of this ratio are given in the caption of Fig.~\ref{fig:EVENTS}.
The theoretical mean of a Landau distribution is undefined due to its 
infinitely long tail.  In reality there is an upper limit
on $\Delta$ and using the measured range to set the limits gives a mean value for 
$\Delta$ that is essentially twice the most probable value, $\Delta_\mathrm{mp}$.
For the DE-20 the mean signal from a individual electron is $\sim$100 times that 
of the readout noise in a given pixel.

In the absence of readout noise the value of $\mathrm{DQE}(0)$ can be calculated from 
the first two moments of $p(\Delta)$ using
\begin{equation}
\mathrm{DQE}(0) = \left( \int p(\Delta) \Delta \, \mathrm{d}\Delta \right)^2  \Big/
\int p(\Delta)\Delta^2 \, \mathrm{d}\Delta .
\label{eqn:MTF_DELTA_DQE}
\end{equation}
If the distributions for the detectors were actually the same 
the detectors would be expected to have the same $\mathrm{DQE}(0)$.  
Using the measured distributions gives apparent values for 
$\mathrm{DQE}(0)$ of 0.34, 0.47 and 0.48 for the DE-20, Falcon~II and K2, respectively.
The value of $\mathrm{DQE}(0)$ calculated using Eqn.~\eqref{eqn:MTF_DELTA_DQE}
is very sensitive to systematic errors in calculating $\Delta$ and
to small differences (especially at high $\Delta$) in $p(\Delta)$. In particular
the value of $\mathrm{DQE}(0)$ for the K2 is much greater than that of the
DE-20 despite the similarity in their distributions as shown in 
Fig.~\ref{fig:EVENTS}d  and the respective
first moments of $p(\Delta)$ being almost identical. The difference arises
from the truncation of $p(\Delta)$ for the K2 at high $\Delta$
relative to what is seen in the other two detectors. The truncation is 
either an artefact of the selected events or indicative of saturation
in the linear output of the K2.

\begin{figure}[t]
\begin{center}
\includegraphics[width=8cm]{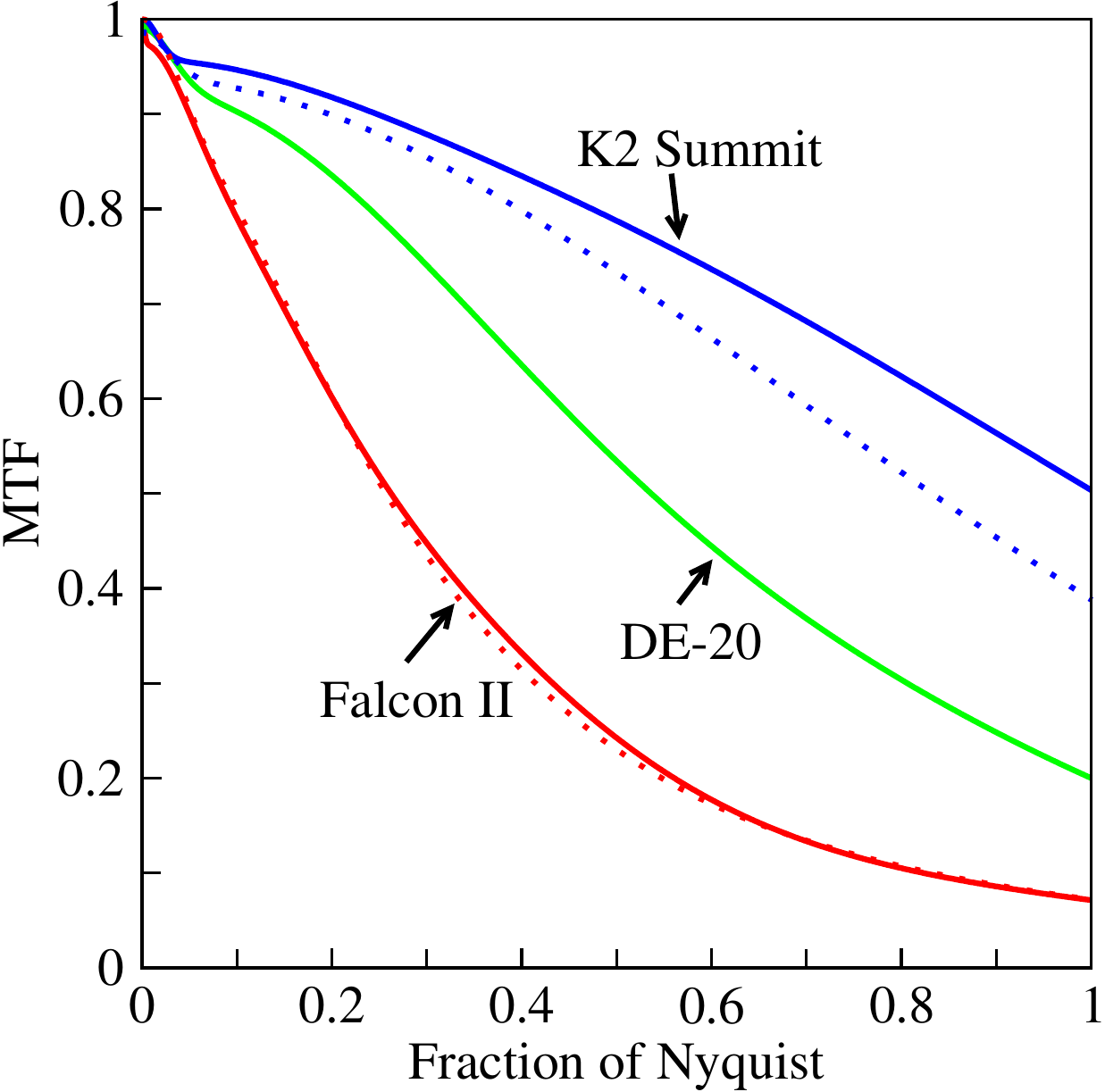}
\end{center}
\caption{
Measured MTF as a function of spatial frequency. The solid lines are for the 
DE-20 (green), Falcon II (red), and K2~Summit in super-resolution mode (blue).
The dotted blue line is the corresponding K2~Summit result in normal resolution mode.
The dotted red line is the MTF obtained for the Falcon~II via the noise power spectra method.
}
\label{fig:MTF}
\end{figure}

\begin{figure}[t]
\begin{center}
\includegraphics[width=7.8cm]{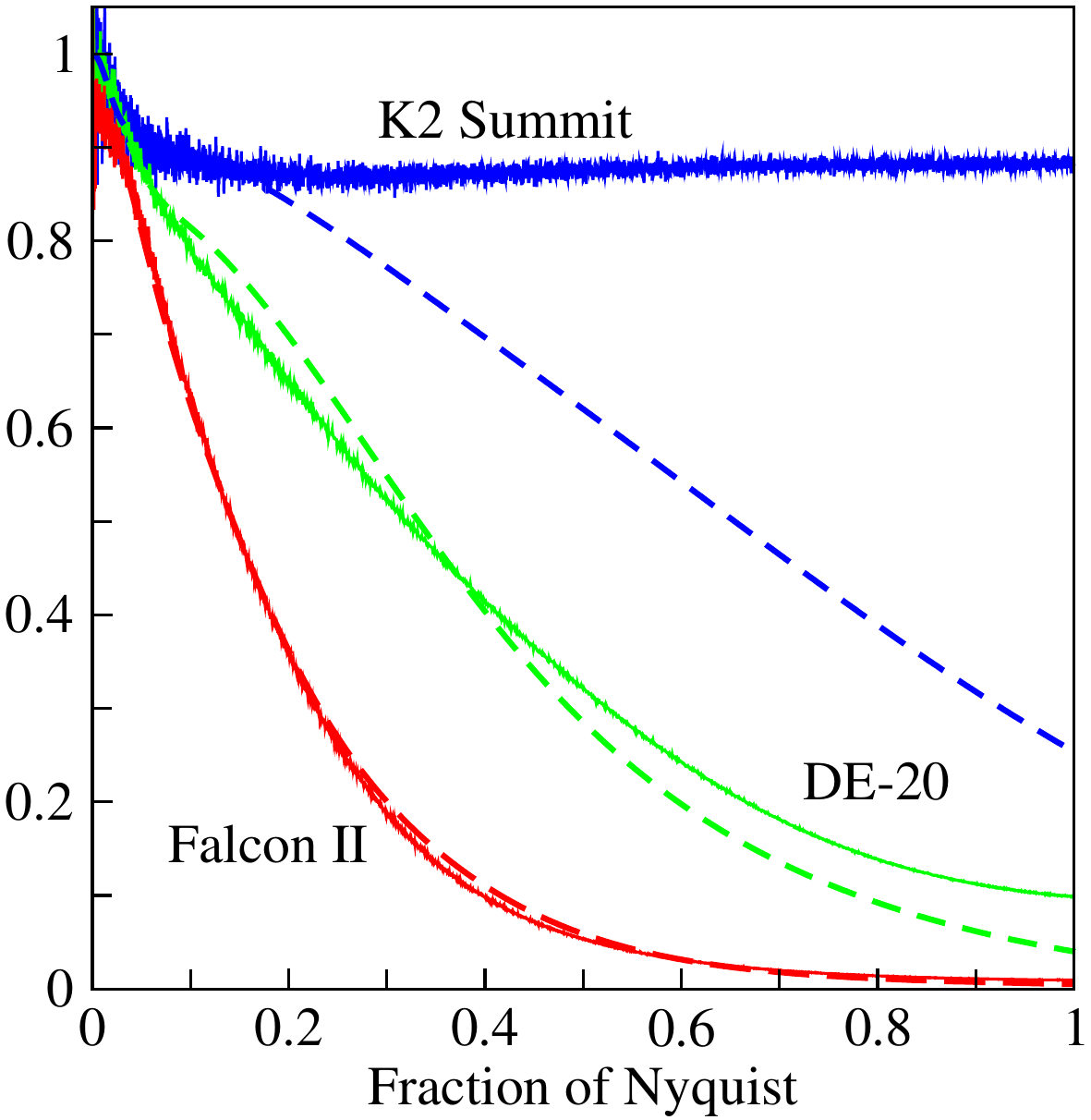}
\end{center}
\caption{
Comparison of the behaviour as a function of spatial frequency 
of  power spectra (solid) and $\mathrm{MTF}^2$ (dashed) for 
the DE-20 (green), Falcon~II (red) and K2~Summit (blue).
The noise power spectra have been scaled to unity at the origin.
}
\label{fig:NOISE}
\end{figure}

The measured MTF as a function of spatial frequency for the
three detectors is shown in Fig.~\ref{fig:MTF}. The 
corresponding edge spread functions and Gaussian expansion fits 
are given as supplementary data.
The MTF of the K2 Summit was obtained both in normal and super-resolution 
(but plotted as a function of the physical Nyquist frequency).
Using super-resolution leads to a significant improvement in the MTF
but the actual enhancement is less than that expected from the 
reduction in the pixel modulation factor, i.e., $ 2\sin(\pi x/2)/\sin(\pi x/4)$
that would be expected if the hardware centroiding algorithm used by the 
K2 worked perfectly.  The results for the Falcon~II MTF calculated 
using both the edge and noise power spectra methods agree very well. 
While the MTF of the Falcon~II
is lowest, the agreement between these two methods for calculating the MTF
indicates that to a first approximation 
its response to incident electrons 
can be described by a simple point spread function.

Fig.~\ref{fig:NOISE} shows  both the
measured noise power spectra for the detectors
(scaled so that they are $\sim$1 at zero spatial frequency) and the 
$\mathrm{MTF}$ results from figure
Fig.~\ref{fig:MTF} plotted as $\mathrm{MTF}^2$. 
The ratio of the $\mathrm{MTF}^2$ to the noise power spectra 
determines the
behaviour of the $\mathrm{DQE}$  
as a function of spatial frequency.
Fig.~\ref{fig:NOISE} illustrates
the contrasting behaviours expected for the DQE in the K2 Summit and Falcon~II
detectors. The noise power spectra of the K2 Summit is essentially constant 
and so the behaviour of the DQE is determined by that of the 
$\mathrm{MTF}^2$.  In contrast the noise power spectra of the Falcon~II and 
the corresponding $\mathrm{MTF}^2$  track each other and so the DQE of the 
Falcon~II will be expected to be relatively constant as a function of 
spatial frequency.

\begin{figure}[!hb]
\begin{center}
\includegraphics[width=8cm]{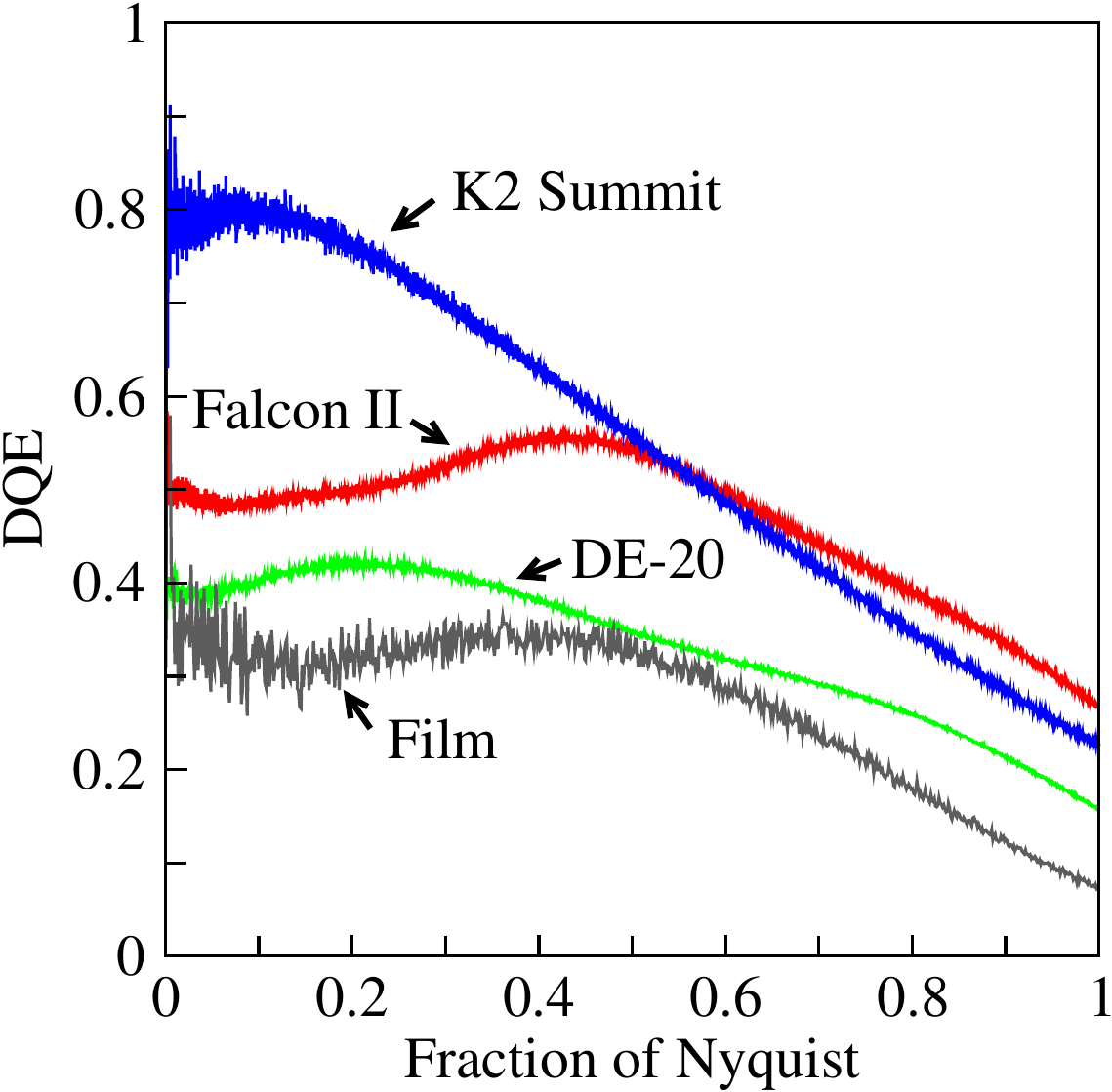}
\end{center}
\caption{
Measured DQE as a function of spatial frequency for the:
DE-20 (green), Falcon~II (red) and 
K2~Summit (blue). The corresponding DQE of photographic
film from \cite{McMullan20091126} is shown in grey.
}
\label{fig:DQE}
\end{figure}

Fig. \ref{fig:DQE} shows the DQE as a function of spatial frequency for 
the three detectors. For reference the corresponding DQE of Kodak SO-163 photographic film 
as given in \cite{McMullan20091126} is also included. 
Clearly the counting mode of the  K2~Summit gives it the  highest
DQE at low spatial frequencies. 
With increasing spatial frequency the DQE of the K2~Summit falls roughly as the MTF$^2$
but those of the DE-20 and Falcon~II stay fairly constant out to
1/2 the Nyquist frequency due to parallel 
falls in their respective noise power spectra. The fall in DQE towards 
the Nyquist frequency is expected due to increasing contributions from aliased
terms in the noise power spectra.
Beyond $\sim$ 3/4 of the Nyquist frequency the DQE of the Falcon~II becomes the
highest of the three detectors. This occurs despite the Falcon~II having the lowest 
MTF (the MTF at the Nyquist frequency of the K2 Summit in super-resolution mode
is nearly 7 times that of the Falcon~II). 

\begin{figure*}[!ht]
\begin{center}
\includegraphics[width=14cm]{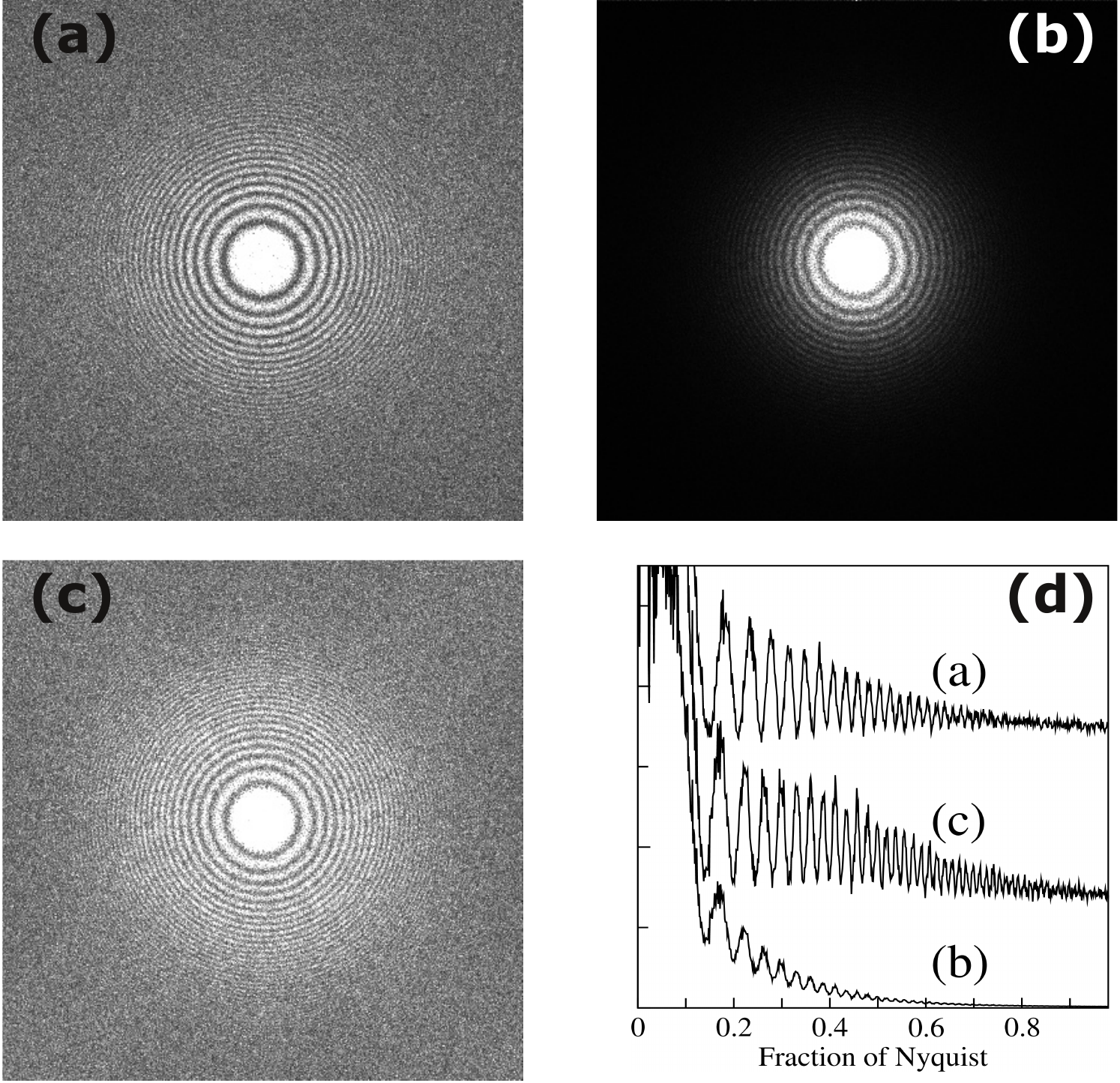}
\end{center}
\caption{
Comparison of the Thon rings seen in the power spectra of images
taken of the same sample using the same number of electrons. 
The power spectra were obtained from a matching 1600x1600 area
from the images.  The power spectra from K2~Summit and Falcon~II 
are shown in (a) and (b) respectively. The results in (c) show
the results of (b) after they have been divided by the 
$\mathrm{NNPS}(\omega)$ as described in the text. The circular
averages over 90 degrees of (a), (b), and (c) are shown in (d).
}
\label{fig:THON}
\end{figure*}

As the Falcon~II and K2~Summit are on the same microscope it 
was possible to compare directly their performance using an image 
of the same sample.  The different pixel sizes and positions of the 
cameras makes it impossible to achieve exactly the same imaging conditions
however the  $\sim$ 2.2\AA\//pixel sampling obtained with the
Falcon~II in TEM mode and nominal magnification setting of
SA37000 was within 5\% of that obtained on the K2~Summit in EFTEM mode 
and nominal magnification setting of SA53000.
The stability of the Krios column allows the same area of a sample to be
seen with essentially the same defocus despite switching back and forth between
EFTEM and TEM modes of the microscope. An area of carbon on a Quantifoil 
grid\footnote{Quantifoil Micro Tools GmBH, www.quantifoil.com} was first pre-irradiated for 5 minutes. A 30 sec exposure at 
3.5 e/pixel/sec of part of this area was recoded in EFTEM mode using 
the K2~Summit and saved in 1 sec blocks.
The microscope was then switched to TEM mode and a long exposure of
essentially the same area taken using the Falcon II in movie mode at an 
exposure rate of 0.66 e/pixel/frame. 
In order to match the exposures on both cameras the last 13 sec of the
K2 Summit image and the first 69 frames from the Falcon image were then
selected. 
The dose rate for the K2~Summit was increased slightly so as to minimise 
drift while a lower
dose rate was used with the Falcon~II to allow the total number
of electrons in each exposure to be closely matched. The exposures from
both detectors were drift corrected and a matching
1600x1600 area from both images selected. The power spectra from 
these are shown in Fig.~\ref{fig:THON}. Despite having a total dose
of only 45.5 electrons per pixel the Thon rings in both images go out
to almost the Nyquist frequency.

The clarity with which the Thon rings in 
Fig.~\ref{fig:THON}a can be seen is partly due to the fact that the
underlying noise power spectra resulting from  the counting mode of 
K2~Summit is essentially flat.  It is not immediately obvious that
the signal to noise ratio of the Thon rings shown in Fig.~\ref{fig:THON}b 
from the Falcon~II image is in fact very similar to that from the K2 at
higher spatial frequency.
In general, subtle variations and small signals, can easily be lost amongst the 
rapidly changing background. 
To avoid this it is useful to mimic the K2~Summit behaviour
and sharpen the images so that the underlying noise is flat. This
requires dividing by a normalised noise power
spectra, $\mathrm{NNPS}(\omega)$ (or equivalently the square of the 
noise transfer function introduced in \citep{MeyerKirkland2000}).
This is illustrated in Fig.~\ref{fig:THON}c where the
$\mathrm{NNPS}(\omega)$ in this case was calculated from the measured MTF
as described in Appendix A. 
The differences between the results for the  K2~Summit in 
Fig.~\ref{fig:THON}a and those of the Falcon~II in 
Fig.~\ref{fig:THON}c are difficult to see and so the 
circularly averaged values are given in 
Fig.~\ref{fig:THON}d.  For display purposes the plots have been
arbitrarily displaced and as 
there was some residual astigmatism in the images the circular average was only carried out
over 90 degrees.  As expected from the DQE results presented in
Fig.~\ref{fig:DQE} there is very little difference between the Thon
ring visibility at high spatial frequencies from the two detectors.

\section{Discussion}

In the study of radiation sensitive samples the 
most important property in a detector is its 
DQE.  The results presented here show 
that there are now three commercially available detectors that have
higher DQE than photographic film.  The low readout noise of
these detectors also means that, unlike film with its finite fog level, 
their DQE remains high even at very low dose rates per image.

The observed differences in the MTF behaviour 
of the detectors will result in cosmetic  differences between 
the resulting images. For example, 
images from the Falcon~II will appear blurred 
relative to those obtained from a high MTF detector such as 
the K2 Summit. But as the MTF is known, the true signal strength 
incident on a detector can be in principle be recovered or  simply
sharpened as in Fig.~\ref{fig:THON}c to better reflect the actual
signal to noise.  Provided a detector has a high DQE the fall in 
signal amplitude with increasing spatial frequency due to poor MTF 
will not effect the resolution that can be reached such as 
with a single particle reconstruction program like RELION\cite{Scheres2012406}.

The counting mode used by the K2~Summit, like that of the Medipix2\cite{McMullan2007401}, 
eliminates readout noise. The 
$\mathrm{DQE}(0)$ is determined by the statistics of, and efficiency in,
counting incident electrons and having no readout noise means that the K2~Summit can
be used to arbitrarily low exposure rates without affecting the DQE.
The readout noise in both the DE-20 and Falcon-II is much lower than the 
average signal left by individual electrons and consequently will only start to 
reduce the value $\mathrm{DQE}(0)$ at very low exposure rates. 
The  value of $\mathrm{DQE}(0)$  in Fig.~\ref{fig:DQE} for 
both the DE-20 and Falcon~II is determined by 
the intrinsic variability in the energy deposited by incident electrons.
Despite having a larger pixel size, the Falcon~II MTF is lower than
that of the DE-20. The Falcon~II is therefore likely to have a thicker
sensitive layer (the epilayer)  so as to  allow
greater diffusion of charge carriers between pixels. Incident 
electrons will on average leave a greater signal in a thicker sensitive
layer and the measured higher DQE of the Falcon~II implies that the
relative variance in this distribution is smaller, though this
difference is difficult to see in Fig.~\ref{fig:EVENTS}d.

By using a counting mode, the K2~Summit is able to escape the intrinsic
variability in the signal deposited by incident high energy electrons and
achieve a higher DQE at low spatial frequency.
Obtaining the maximum performance from a counting detector 
requires minimising coincidence losses
and despite the K2 Summit running at 400~fps the coincidence losses 
quickly mount up. At 1~e/pixel/sec the loss is negligible but 
at 10 e/pixel/sec the results in \cite{Ruskin2013385} and the 
analysis in Appendix B say that the initial DQE(0) will be down by 
over 12\%. Obtaining maximum performance from the K2~Summit requires
using longer exposures that will inevitably need some form of stage
drift correction.

The extrapolated value of DQE(0) in
Fig.~\ref{fig:DQE} for the K2~Summit is much lower than would be expected purely from 
coincidence losses at 1.1 e/pixel/sec.
The many problems faced by counting detectors are illustrated in Fig.~\ref{fig:EVENTS}.
Some single events are spread over several pixels while others resemble tracks that the 
counting algorithm can either reject, treat as a single  event, or treat as multiple events. 
The distribution
of event energies, as shown in Fig.~\ref{fig:EVENTS}d,  means that some events 
will always be barely above the readout noise and for practical reasons the loss 
of a proportion of these has to be accepted.  
Despite careful dark and gain calibration there will also inevitably be a few "hot" pixels 
that produce much higher noise levels.  
For integrating detectors,
such as the DE-20 or Falcon, this is unimportant as the signal due to the
readout noise is only a small fraction of the average signal from one electron. 
With a counting detector 
false counts from the noise have the same weight as incident 
electrons. If a threshold is set so as to avoid generating counts in these pixels
then too many true events will be lost and for this reason images from counting detectors
often contain erroneously high counts in some pixels. In \cite{McMullan20091411} the
so called "hot" pixels were identified as their signal was only in one pixel while
true signals from incident electrons always spread into neighbouring pixels.

The spatial frequency dependence of the DQE in electron counting implementation used in 
the K2~Summit is determined primarily by that of the $\mathrm{MTF}^2$. This is a consequence
of the almost flat noise power spectra resulting from putting the signal from an electron
into a single pixel.  Getting the best performance from the K2~Summit therefore requires 
maximising the MTF and so using the super-resolution mode.  The counting implementation 
as used in the K2~Summit is not the only way to process individual incident electron 
events in order to build up an image.  In particular, an improved DQE at higher spatial 
frequency can  be obtained 
through retaining more sub-pixel information by 
using a distribution with finite spatial extent centred on the inferred incident position
instead of putting all the weight for an electron in a single pixel.
The optimal method of processing electron event images  will of course depend on the incident 
electron energy and the  underlying properties of the detector. For example in \cite{McMullan20091411} it 
was found that a better DQE at higher spatial frequencies could be obtained by 
treating an incident electron image as a probability distribution 
for the incident position of an electron on the
detector.  This approach worked well for that particular combination of detector and incident
electron energy but does not necessarily work in general. In particular this 
approach is not suited to the case such as the Falcon~II where the event distribution is
dominated by the effects of carrier diffusion.  In
\cite{McMullanJINST2011} it was argued that instead of the raw image being used as a 
probability distribution the image should first be processed to remove effects such 
as a known point spread function and the processed image then be used
as a probability distribution.

The lack of readout noise and high DQE at lower spatial frequency means that
K2~Summit is  particularly well suited to use in tomography.
The near linear increase in DQE of the K2~Summit
with decreasing spatial frequency means that there is always an
advantage in improved signal-to-noise from using
higher magnification. This also results in a higher dose rate on the sample
and so shorter exposures but at the price of a reduced field of view.

The DQE of both the DE-20 and Falcon~II detectors is relatively flat between zero 
and 50\% of the Nyquist frequency. Because of this there is no advantage 
with these detectors in binning pixels or going to higher magnification.  
As lower spatial frequency information in a sample tends to be more resistant to radiation damage,
higher signal to noise for these frequencies can be obtained using higher dose. To
make use of this and  yet avoid degrading  higher spatial frequency information,  images must be
acquired as movies with the spatial frequency information from the frames
weighted by a dose dependent weighting such as that used in \cite{BetaGal2014}.
In CryoEM  studies where particles that can be aligned  with relatively low spatial frequency 
information from the sample ($\sim 20 \AA$) 
there should therefore be very little difference between the resolution that can be 
attained with all three of the detectors. Higher DQE will
show up in terms of a reduction in the number of particles required to reach 
a given resolution. The lower DQE of the DE-20 will in part be compensated for 
by decreased number of images needed with its greater field of view. The higher dose rate
that can be used on the DE-20 also leads to shorter overall exposure times which is 
advantageous on microscopes with side-entry cold stages that are inherently
less stable.  For particles whose images cannot be easily oriented, such as smaller or featureless
particles, there is however no substitute for the higher DQE available by
using higher magnification with the K2~Summit.

While all the detectors studied here represent an improvement in both  DQE and
convenience of use over film, in the long run the combination 
of high DQE, low readout noise and ability to capture time series data may produce
the greatest impact.  For example it is now clear that one
of the major reasons for the low amplitude in cryoEM of high 
spatial frequency information is not just radiation damage but the 
presence of substantial initial movement of  samples as the beam is 
turned on. The ability provided by the new detectors to see and quickly diagnose 
this movement will hopefully lead to a way of reducing, or eliminating,  it 
and so lead to a further leap in ease of use and resolution that 
can be obtained with cryoEM.

%
%

\section*{Acknowledgements and declaration of interest} 
We are grateful to Helen Saibil and Elena Orlova at Birkbeck College for 
their support in carrying out this comparative analysis.
Three of the authors (McMullan, Faruqi \& Henderson) were part of a 
consortium including FEI that explored various CMOS detector designs
for use with electron microscopy but the actual details of the 
Falcon camera 
design were not part of that work.

\section*{Appendix A: Noise power spectra and MTF}
The MTF at spatial frequency, $\omega$,  measured in terms of the 
Nyquist frequency is can be 
expressed as \citep{McMullan20091126}
\begin{equation}
\mathrm{MTF}(\omega) = \mathrm{sinc}( \pi \omega/2 )\,  \mathrm{MTF}_0(\omega) 
\label{eqn:MTF_CIRCULAR}
\end{equation}
in which $\mathrm{sinc}(\pi \omega/2) = \sin(\pi\omega/2)/(\pi\omega/2)$  is 
the pixel modulation factor and 
$\mathrm{MTF}_0(\omega)$ can be viewed as 
the Fourier transform of an
intrinsic point spread function, $\mathrm{PSF}(r)$ of the detector. 
For convenience $\mathrm{MTF}_0(\omega)$ can be expanded as
normalised sum of 
Gaussian functions with different length scales, $\lambda_k$, i.e., 
\begin{equation}
\mathrm{MTF}_0(\omega) = \sum_k a_k \exp( -\pi^2\lambda_k^2 \omega^2/4 ) 
\label{eqn:MTF0}
\end{equation}
in which the weights, $a_k$, obey  $\sum_k a_k = 1$ and the values of
$\lambda_k$, and $a_k$ are chosen to fit the 
measured edge spread function.  
The corresponding intrinsic point spread function 
\begin{equation}
\mathrm{PSF}_0( r) = \sum_k a_k \exp( -r^2/\lambda_k^2 )/\pi\lambda_k^2
\label{eqn:PSF}
\end{equation}

The noise power spectrum, $\mathrm{NPS}(\omega)$, of
$n$ randomly distributed  electrons with response given by Eqn.~\eqref{eqn:PSF} 
is proportional to the modulus squared of the 
Fourier transform of this response.
The Fourier transform of Eqn.~\eqref{eqn:PSF} extends
beyond the Nyquist frequency and in the  
$\mathrm{NPS}(\omega)$ these will appear as aliased terms\cite{MeyerKirkland2000}.
Beyond 50\% of the Nyquist frequency, non-circular terms need to be 
included the noise power spectra so that it becomes
$\mathrm{NPS}(\omega_x, \omega_y )$ in which 
$\omega_x$ and $\omega_y$ are the fractions of the Nyquist frequency in the 
$x$ and $y$ directions.  In the simplest case the additional non-circular terms 
arise from the pixel modulation term so that
Eqn.~\eqref{eqn:MTF_CIRCULAR} becomes
\begin{equation}
\mathrm{MTF}(\omega_x, \omega_y )  = 
\hbox{sinc}(\pi \omega_{x} /2) \, \hbox{sinc}(  \pi \omega_y /2 )  \,
\mathrm{MTF}_0\left(\sqrt{ \omega_x^2 + \omega_y^2 } \right).
\end{equation}
and
\begin{equation}
\mathrm{NNPS}(\omega_x, \omega_y ) =    
\sum_{i,j}  
\hbox{MTF}^2( \omega_{xi}, \omega_{yj})
\end{equation}
where the sum is over all aliased terms $(\omega_{xi},\omega_{yj}) = (\omega_x + i, \omega_y + j )$
with $i,j = 0, \pm 2, \pm 4, \dots$  (in units of the Nyquist frequency). 
The results in Fig.~\ref{fig:THON}c  for the Falcon~II were obtained 
by dividing the measured noise power spectra by the  corresponding
$\mathrm{NNPS}(\omega_x, \omega_y )$ calculated from the 
parameters $a_k$ and $\lambda_a$ used to fit the MTF.

\section*{Appendix B: Counting detector probabilities}
For a perfect counting detector in which there are 
a small number, $n$, of 
electrons incident per pixel in a frame,  the probability of a single event, $p_1$,
is given by the product of the probability one or more electrons landing on a pixel 
multiplied by the probability of no electrons in the surrounding 8 pixels. 
Similarly the probability of a double event, $p_2$, is the probability of one or 
more electrons landing on two adjacent pixels and no electrons landing on the 
surrounding pixels. There are four possible ways to 
have two  neighbours and these have either 10 or 12 surrounding pixels depending whether the 
neighbours are vertical/horizontal or diagonal. Similarly for a 3 pixel event, $p_3$,
but now there are 20 ways of for this to occur. The probabilities for the events
are given by
\begin{eqnarray}
p_1 &=&  e^{-8n}\, (1-e^{-n})   \\
p_2 &=&  (2e^{-10n} + 2e^{-12n} )\,(1-e^{-n})^2 \\
p_3 &=&  (6e^{-12n} + 8e^{-14n} + 4e^{-15n} + 2e^{-16n} )\,( 1- e^{-n})^3.
\end{eqnarray} 
For small $n$, the probability of recording an event is 
$p_t \approx p_1 + p_2 + p_3$ so that 
\begin{equation}
p_t = n - \frac{9}{2}n^2 + \frac{49}{6}n^3 + \ldots .
\label{eqn:PT}
\end{equation}
The counting algorithm used by in the K2~Summit leads to a very simple auto-correlation
function with a value equal to $p_t$ at the origin, a value of zero in pixels surrounding the origin
and a value of $p_t^2$ everywhere else. From this the expected power spectra can be
calculated.  In particular, in units where the power spectra for $n$ random electrons
is given as $n$, the power spectra at the origin (0,0),  Nyquist frequency (1,0) 
and the corner (1,1) are equal to $p_t - 9p_t^2$, $p_t+3p_t^2$ and $p_t-p_t^2$ 
respectively.  Note that the output noise at zero spatial frequency,
given by $p_t-9p_t^2$,  is less than the corresponding input noise of $n$.
This reduced noise at zero spatial frequency does not lead to an increased DQE 
because the output signal is also reduce by the gain multiplied by the MTF. 
At zero spatial frequency this is given
by $\mathrm{d} p_t/\mathrm{d} n $,
and is correspondingly reduced so that 
\begin{equation}
\hbox{DQE}_n(0) = \left( \frac{ \mathrm{d} p_t}{\mathrm{d} n}\right)^2\frac{n}{p_t - 9p_t^2} \approx  1 - \frac{9}{2}n - \frac{239}{12}n^2.
\label{eqn:DQEn}
\end{equation} 
As a result if the K2~Summit is operated at 10 e/pixel/sec, i.e,  
$n = 0.025$,   Eqn.~\ref{eqn:PT}
gives a 12\% coincidence loss  and 
Eqn.~\ref{eqn:DQEn}  predicts a 14\% drop in the $\mathrm{DQE}(0)$.



\bibliography{detectors}

\begin{thebibliography}{10}
\expandafter\ifx\csname url\endcsname\relax
  \def\url#1{\texttt{#1}}\fi
\expandafter\ifx\csname urlprefix\endcsname\relax\def\urlprefix{URL }\fi
\expandafter\ifx\csname href\endcsname\relax
  \def\href#1#2{#2} \def\path#1{#1}\fi

\bibitem{DaintyShaw}
J.~Dainty, R.~Shaw, Image Science, Academic Press, New York, 1974.

\bibitem{Faruqi2005}
A.~R. Faruqi, R.~Henderson, M.~Prydderch, R.~Turchetta, A.~P., A.~Evans,
  {M}{A}{P}{S} detector, Nucl. Instr. and Meth. 11 (2005) 2001--2009.

\bibitem{Milazzo2005152}
A.-C. Milazzo, P.~Leblanc, F.~Duttweiler, L.~Jin, J.~C. Bouwer, S.~Peltier,
  M.~Ellisman, F.~Bieser, H.~S. Matis, H.~Wieman, P.~Denes, S.~Kleinfelder,
  N.-H. Xuong, Active pixel sensor array as a detector for electron microscopy,
  Ultramicroscopy 104~(2) (2005) 152 -- 159.

\bibitem{McMullan20091126}
G.~McMullan, S.~Chen, R.~Henderson, A.~Faruqi, Detective quantum efficiency of
  electron area detectors in electron microscopy, Ultramicroscopy 109~(9)
  (2009) 1126 -- 1143.
\newblock \href
  {http://dx.doi.org/http://dx.doi.org/10.1016/j.ultramic.2009.04.002}
  {\path{doi:http://dx.doi.org/10.1016/j.ultramic.2009.04.002}}.

\bibitem{McMullan20091411}
G.~McMullan, A.~Clark, R.~Turchetta, A.~Faruqi, Enhanced imaging in low dose
  electron microscopy using electron counting, Ultramicroscopy 109~(12) (2009)
  1411 -- 1416.
\newblock \href
  {http://dx.doi.org/http://dx.doi.org/10.1016/j.ultramic.2009.07.004}
  {\path{doi:http://dx.doi.org/10.1016/j.ultramic.2009.07.004}}.

\bibitem{MeyerKirkland1998}
R.~R. Meyer, A.~Kirkland, The effects of electron and photon scattering on
  signal and noise transfer properties of scintillators in {C}{C}{D} cameras
  used for electron detection, Ultramicroscopy 75~(1) (1998) 23--33.

\bibitem{Boothroyd201318}
C.~Boothroyd, T.~Kasama, R.~Dunin-Borkowski, Comparison of approaches and
  artefacts in the measurement of detector modulation transfer functions,
  Ultramicroscopy 129~(0) (2013) 18 -- 29.
\newblock \href
  {http://dx.doi.org/http://dx.doi.org/10.1016/j.ultramic.2013.03.001}
  {\path{doi:http://dx.doi.org/10.1016/j.ultramic.2013.03.001}}.

\bibitem{Landau1944}
L.~Landau, On the energy loss of fast particles by ionization, J.Phys. USSR 8
  (1944) 201 -- 205.

\bibitem{MeyerKirkland2000}
R.~R. Meyer, A.~I. Kirkland, Characterisation of the signal and noise transfer
  of {C}{C}{D} cameras for electron detection, Microscopy Research and
  Technique 49~(3) (2000) 269--280.

\bibitem{Scheres2012406}
S.~H. Scheres, A {B}ayesian view on cryo-em structure determination, Journal of
  Molecular Biology 415~(2) (2012) 406 -- 418.

\bibitem{McMullan2007401}
G.~McMullan, D.~Cattermole, S.~Chen, R.~Henderson, X.~Llopart, C.~Summerfield,
  L.~Tlustos, A.~Faruqi, Electron imaging with medipix2 hybrid pixel detector,
  Ultramicroscopy 107~(4–5) (2007) 401 -- 413.
\newblock \href
  {http://dx.doi.org/http://dx.doi.org/10.1016/j.ultramic.2006.10.005}
  {\path{doi:http://dx.doi.org/10.1016/j.ultramic.2006.10.005}}.

\bibitem{Ruskin2013385}
R.~S. Ruskin, Z.~Yu, N.~Grigorieff, Quantitative characterization of electron
  detectors for transmission el ectron microscopy, Journal of Structural
  Biology 184~(3) (2013) 385 -- 393.

\bibitem{McMullanJINST2011}
G.~McMullan, R.~Turchetta, A.~R. Faruqi, Single event imaging for electron
  microscopy using maps detectors, Journal of Instrumentation 6~(04) (2011)
  C04001.

\bibitem{BetaGal2014}
K.~R. Vinothkumar, G.~McMullan, R.~Henderson, Molecular mechanism of
  antibody-mediated activation of $\beta$-galactosidase, Structure 22 (2014)
  621--627.
\newblock \href {http://dx.doi.org/10.1016/j.str.2014.01.011}
  {\path{doi:10.1016/j.str.2014.01.011}}.

\end{thebibliography}

\end{document}